\documentclass[useAMS,usenatbib]{mn2e}

\usepackage{psfig,epsfig,supertabular,wrapfig,placeins}
\usepackage{graphicx,amsmath,amssymb,subfigure,multirow,units}
\usepackage{marvosym,nicefrac}
\usepackage{graphics}
\usepackage{natbib}
\usepackage{amssymb}
\usepackage{amsmath}
\usepackage{euscript}
\usepackage{setspace}
\usepackage{fancyhdr}
\usepackage{afterpage,rotating,url}
\usepackage{times}

\newcommand{\etal}{{et al.}}                   

\newcommand{\Msolar}{\mbox{\,$\rm M_{\odot}$}}        
\newcommand{\Lsolar}{\mbox{\,$\rm L_{\odot}$}}        
 
\newcommand{\asec}{\ensuremath{^{\prime\prime}}}
\newcommand{\amin}{\ensuremath{^{\prime}}}

\newcommand{\Heii}{He{\sc ii}}

\newcommand{\gtsim}{\mbox{{\raisebox{-0.4ex}{$\stackrel{>}{{\scriptstyle\sim}}$}
}}}

\title[Infrared \& millimetre-wavelength evidence for Cold
  Accretion]{Infrared and millimetre-wavelength evidence for cold accretion within a $z=2.83$ Lyman-$\alpha$ Blob}

\author[D.J.B.~Smith, M.J.~Jarvis, M.~Lacy \& A.~Mart\'inez-Sansigre]{Daniel J.B. Smith$^{1,2}$\thanks{E-mail:
    djs@astro.livjm.ac.uk(DS)}, Matt J. Jarvis$^{3}$, Mark Lacy$^{4}$ and Alejo Mart\'inez-Sansigre$^{5}$ \\
    $^{1}$Astrophysics Research Institute, Liverpool John Moores University, Twelve Quays House, Egerton Wharf, Birkenhead,\\ CH41 1LD, UK\\ 
    $^{2}$Department of Astrophysics, University of Oxford, Denys Wilkinson Building, Keble Road, Oxford, OX1 3RH, UK \\
    $^{3}$Centre for Astrophysics, Science \& Technology Research Institute, University of Hertfordshire, Hatfield, Herts, AL10 9AB, UK \\
    $^{4}$Spitzer Science Center, California Institute of Technology, Pasadena, CA, USA\\
    $^{5}$Max-Planck-Institut f\"ur Astronomie, K\"onigstuhl 17, D-69117 Heidelberg, Germany}
    
\begin{document}

\date{\today}

\pagerange{\pageref{firstpage}--\pageref{lastpage}} \pubyear{2002}

\maketitle

\label{firstpage}

\begin{abstract}
This paper discusses infrared and millimetre-wavelength observations
of a Lyman-$\alpha$ blob discovered by Smith \& Jarvis, a candidate
for ionization by the cold accretion scenario discussed in Fardal et
al. and Dijkstra et al. We have observed the counterpart galaxy at
infrared wavelengths in deep observations with the {\it Spitzer Space
  Telescope} using the IRAC 3.6, 4.5, 5.8 \& 8.0$\mu$m and MIPS
24$\mu$m bands, as well as using the Max-Planck Millimeter Bolometer
Array at a wavelength of 1.2mm with the IRAM 30 metre telescope. These
observations probe the $\gtsim95$kpc Lyman-$\alpha$ halo for the
presence of obscured AGN components or the presence of a violent
period of star formation invoked by other models of ionisation for
these mysterious objects. 24$\mu$m observations suggest that an
obscured AGN would be insufficiently luminous to ionize the halo, and
that the star formation rate within the halo may be as low as $<$140
\Msolar yr$^{-1}$ depending on the model SED used. This is reinforced
by our observations at 1.2mm using MAMBO-2, which yield an upper limit
of SFR $<$550\Msolar yr$^{-1}$ from our non-detection to a 3$\sigma$
flux limit of 0.86 mJy beam$^{-1}$.  Finding no evidence for either
AGN or extensive star formation, we conclude that this halo is ionised
by a cold accretion process. We derive model SEDs for the host galaxy,
and use the Bruzual \& Charlot and Maraston libraries to show that the
galaxy is well described by composite stellar populations of total
mass 3.42 $\pm\ 0.13 \times 10^{11}$\Msolar\ or 4.35 $\pm\ 0.16 \times
10^{11}$\Msolar\ depending on the model SEDs used.

\end{abstract}


\section{Introduction}


Lyman-$\alpha$ Blobs, discovered by Steidel \etal\ (2000), consist of
amoebic structures emitting profusely at the rest-frame wavelength of
the Lyman-$\alpha$ emission line, 1216\AA. Whilst these very large
Lyman-$\alpha$ emitting haloes are reminiscent of the extended
emission line regions observed around powerful high redshift radio
galaxies (e.g. Villar-Mart\'in \etal, 2007b), they typically have less
than 1\% of the associated radio flux, raising the question as to what
ionises the neutral Hydrogen in order to enable the emission of
Lyman-$\alpha$ photons. Three of the most plausible explanations for
this are:

\begin{itemize}
\item LABs contain hidden QSOs (e.g. Haiman \& Rees 2001; Weidinger,
  M\o ller \& Fynbo 2004; Weidinger et al. 2005; Barrio \etal,
  2008). The luminous nature of LABs with typically $L = 10^{44}$ erg
  s$^{-1}$ in the Lyman-$\alpha$ emission line alone suggests that the
  hard spectra, and bolometric luminosity of QSOs are prime candidates
  to power such a galaxy's emission.
\item The Lyman-$\alpha$ emission comes from a dust-enshrouded,
  extreme starburst galaxy with a large scale superwind due to large
  numbers of luminous and short-lived OB stars (e.g. Taniguchi \&
  Shioya 2000; Ohyama \etal ~2003; Wilman \etal ~2005, Matsuda
  \etal\ 2007). Observational evidence from e.g. Chapman \etal\ (2001)
  and Geach \etal\ (2007) associating LABs with sub-millimetre
  galaxies suggests that some LABs contain enshrouded starbursts
  forming stars at rates of $\sim 1000$\Msolar yr$^{-1}$.
\item We are observing the cooling radiation of a collapsing
  proto-galaxy inside a dark matter halo's gravitational potential
  (the so-called ``cold accretion'' model - e.g. Haiman, Spaans \&
  Quataert, 2000, Fardal \etal, 2001, Matsuda \etal, 2004, Dijkstra
  \etal\ 2006a,b, Nilsson \etal, 2006, Dijkstra \etal, 2007, Smith \&
  Jarvis, 2007).  A cold accretion scenario invokes collisional
  excitation and re-radiation of accreting neutral gas to power the
  profuse Lyman-$\alpha$ emission extended over the halo.
\end{itemize}

The sources of ionisation residing within these galaxies (which are
sometimes extended over many tens of kiloparsecs), are currently
thought to be diverse. This result is due to the discovery of a 200kpc
LAB in Dey \etal (2005) associated with a 24$\mu$m detection
attributed to an obscured AGN, to the discovery by Chapman
\etal\ (2001) of a highly luminous sub-millimetre source (with $L_{\rm
  bol} \sim 10^{13}$\Lsolar) indicative of a very high star formation
rate ($\sim 1000 \Msolar$ yr$^{-1}$) residing within LAB1 from Steidel
\etal\ (2000), and to the non-detection of any apparent source of
ionization associated with a LAB in the GOODS-South field by Nilsson
\etal\ (2006).

\begin{figure}
\centering
\includegraphics[height=0.90\columnwidth,angle=270]{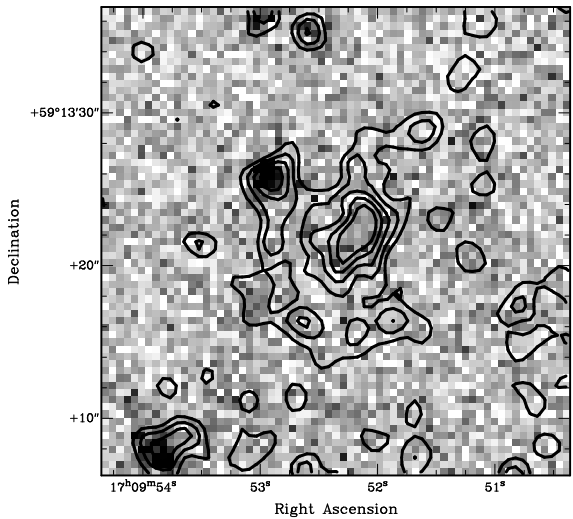}
\caption{Narrow-band contour map overlaid on the Sloan-g' band
  data. This frame is $\sim$33\asec on a side and North is up while
  East is to the left. The object to the North and East of the core of
  the Lyman-$\alpha$ emission is a low-redshift interloper, identified
  as being at $z=0.84$ due to the presence of [O{\sc ii}]$_{3727}$,
  [Ne{\sc iv}]$_{2424}$, Mg{\sc ii}$_{2799}$, and [C{\sc ii}]$_{2326}$
  emission in our optical spectroscopy (for more details see Smith \&
  Jarvis, 2007).}
\label{blob_contour}
\end{figure}

Smith \& Jarvis (2007) discovered only the second known Lyman-$\alpha$
Blob thought to be ionized by the process of cold accretion, residing
at $z = 2.83$. In figure \ref{blob_contour} we show the Sloan-g' band
image of the LAB from our survey, overlaid with a contour map of the
\Heii\ narrow-band data, sensitive to Lyman-$\alpha$ emission at the
redshift of the LAB. This LAB was found to be extended over at least
$\gtsim 95$kpc, with a Lyman-$\alpha$ luminosity of $L_{Ly\alpha} =
2.1 \pm 0.5 \times 10^{43}$ erg s$^{-1}$.

The LAB in question is located at 17$^h$09$^m$52.3$^s$
+59$^\circ$13$^\prime$ 21.72\asec\ (J2000), on the very edge of the
Extragalactic component of the {\it Spitzer First Look Survey}
(Marleau \etal, 2004, Fadda \etal, 2006), and was covered with any
S\slash N at all only in the 4.5$\mu$m band (IRAC channel 2 - Lacy et
al., 2005). In any case, the FLS observations were not deep enough for
the type of study proposed here (see below).

AGN themselves are expected to be particularly dusty due to the torus
invoked by schemes of AGN unification (Antonucci, 1993) to explain the
different observed species; this also makes them bright at
mid-infrared wavelengths since the warm torus is thought to reprocess
the X-ray and UV photons emitted by the central engine. Indeed,
through their mid-infrared emission, powerful AGN (quasars) can be
identified up to high redshifts (e.g. Lacy et al., 2004,
Mart\'inez-Sansigre \etal, 2005), even when they are so heavily
obscured that they are undetected at X-ray energies (e.g. Polletta
\etal, 2006, Lacy \etal, 2007, Mart\'inez-Sansigre \etal, 2007).

In the event that there is a 24$\mu$m detection residing within the
LAB, through studying its IR SED we would be able to distinguish
between an obscured AGN and the starburst SED that would be expected
if there was extensive ongoing star formation in the LAB counterpart
galaxy.

LABs ionized by a starburst are found to have very high star formation
rates based on their fluxes at sub-millimetre wavelengths equivalent
to $\sim 1000$\Msolar\ yr$^{-1}$ (e.g. Chapman \etal\ 2001, Geach
\etal, 2007). The recent high-resolution Sub-Millimetre Array (SMA)
observations of Matsuda \etal\ (2007) did not detect LAB1 from Steidel
(2000), despite the bright sub-millimetre continuum measured by
Chapman \etal\ (2001). Matsuda \etal\ (2007) argued that the
sub-millimetre continuum was most likely resolved out by the high
spatial resolution interferometric SMA observations, suggesting that
the spatial extent of the sub-millimetre emitting region was
$\gtsim$30kpc, comparable to the extent of the Lyman-$\alpha$ emission
itself. This reinforces the super-wind model for this LAB, in which
rapid star formation within the halo is widely distributed, and could
be powering the Lyman-$\alpha$ emission. Observing this new halo from
Smith \& Jarvis (2007) at millimetre wavelengths then provides an
additional constraint on the properties of the ionizing source
residing within.

Here we present the results of two independent tests for the presence
of starbursting or AGN components enshrouded within the
structure. This paper is organised as follows; in section
\ref{observations} we describe our {\it Spitzer space telescope} and
MAMBO-2 observations, while in section \ref{sec:Results} we present
our results, including our improved constraints on this galaxy's
spectral energy distribution (SED), and section \ref{conclusions}
presents our conclusions. Throughout this paper the AB magnitude
system is used (Oke \& Gunn, 1983), and a standard cosmology is
assumed in which $H_{0}$ = 71 km s$^{-1}$, $\Omega_{M}$ = 0.27 and
$\Omega_{\Lambda}$ = 0.73 (Dunkley \etal, 2008).

\section{Observations}
\label{observations}

\subsection{Spitzer Space Telescope Observations}
\label{sec:midIR_obs}


We observed the region centred on this LAB in 21 dithered 30s
exposures in IRAC channels 1 \& 3, and 33 $\times$ 30s in IRAC
channels 2 \& 4 (10 \nicefrac{1}{2} and 16 \nicefrac{1}{2} minutes in
total, respectively). We designed our 24$\mu$m MIPS observations to
use a seven-point jittering positional offset algorithm with offsets
corresponding to the six nodes of a hexagon, plus the central
position. We observed at each position for 501s, resulting in a total
of 3507s on source. The jittering pattern was employed to ensure
effective chip artefact rejection, and result in a reasonably uniform
depth over the central 5\amin\ of the field of view. No new MIPS
channel 2 \& 3 data were acquired.

\begin{figure*}
\centering
\subfigure[Channel 1 - 3.6$\mu$m]{\includegraphics[height=0.90\columnwidth, angle=-90]{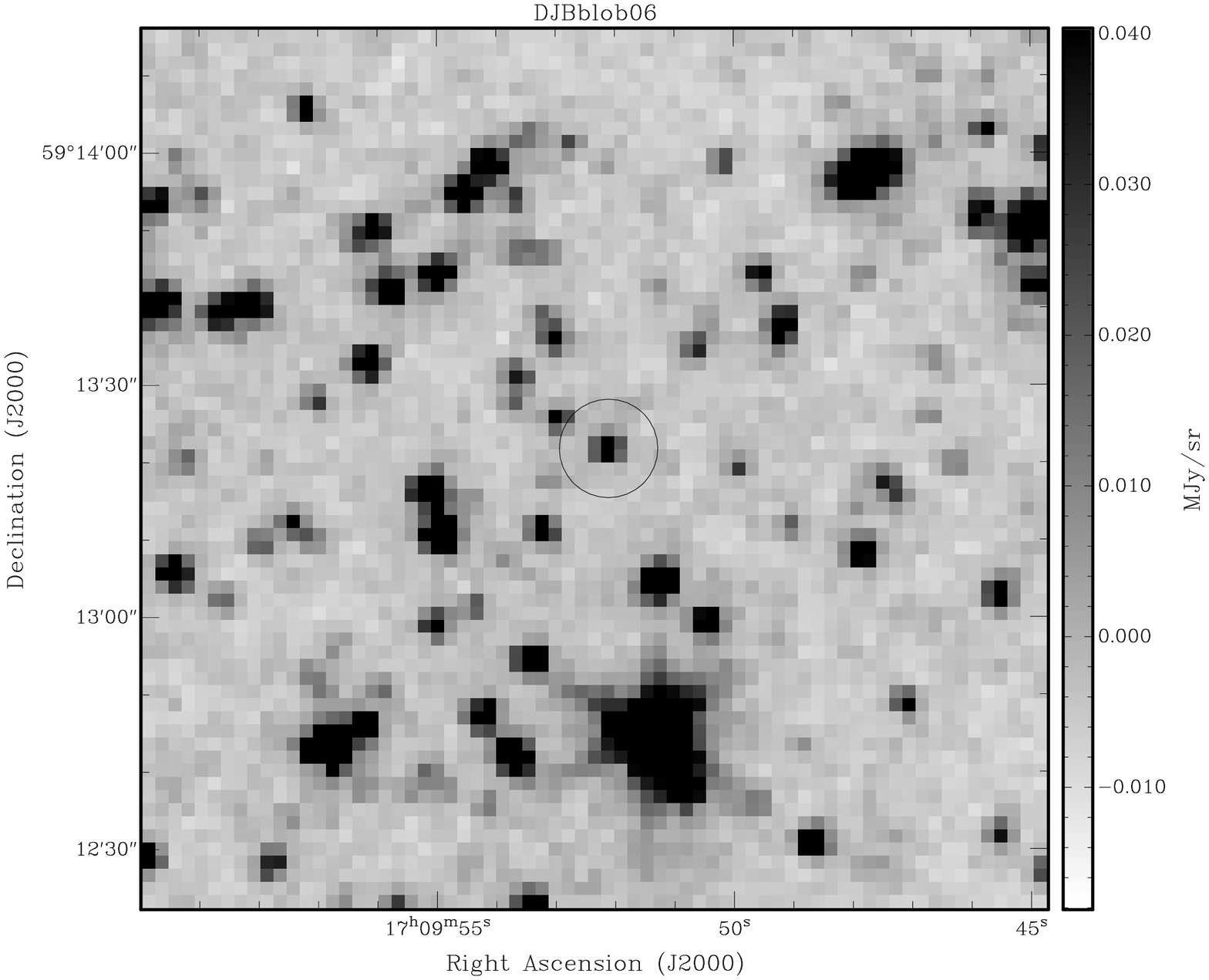}}
\subfigure[Channel 2 - 4.5$\mu$m]{\includegraphics[height=0.90\columnwidth, angle=-90]{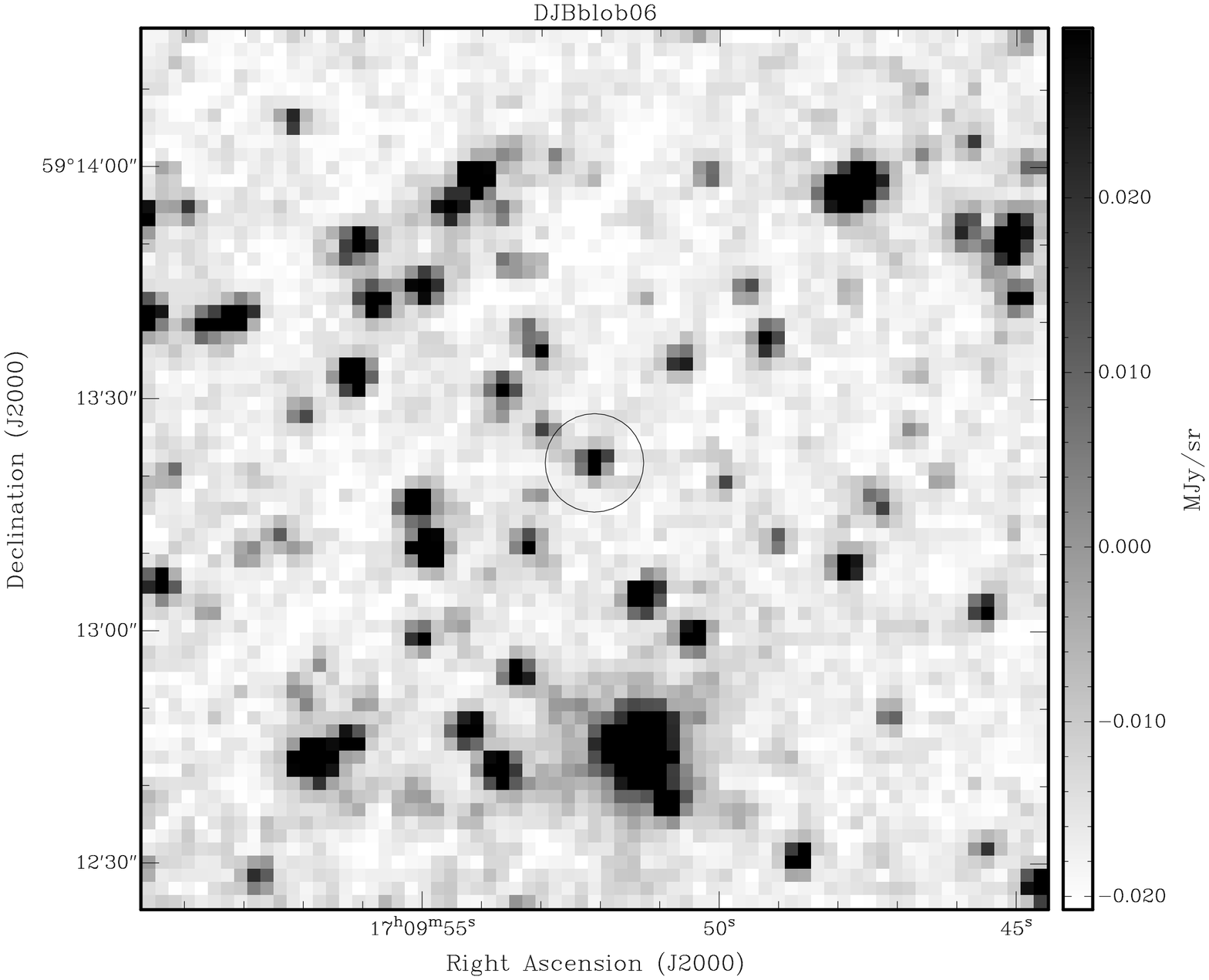}}
\subfigure[Channel 3 - 5.8$\mu$m]{\includegraphics[height=0.90\columnwidth, angle=-90]{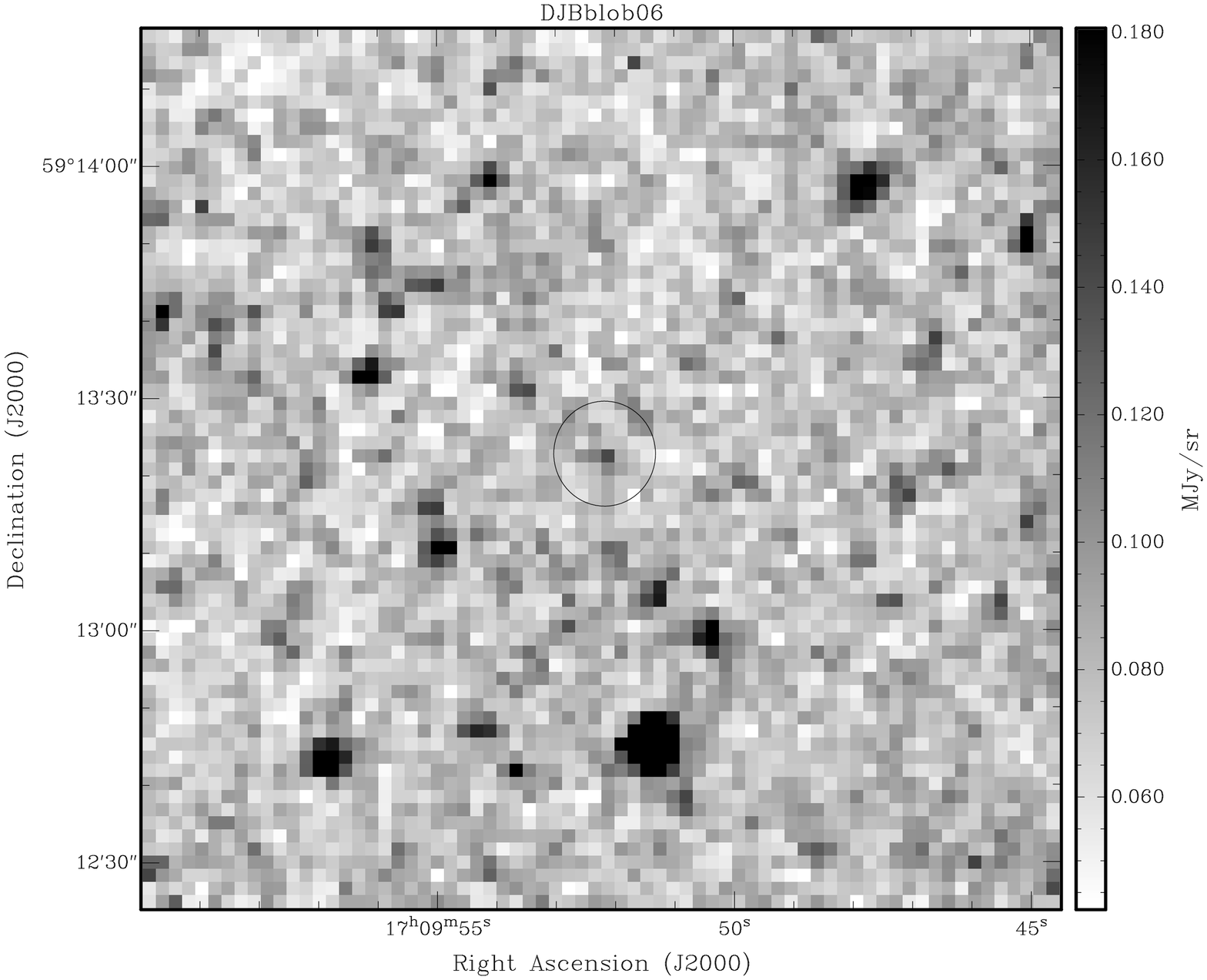}}
\subfigure[Channel 4 - 8.0$\mu$m]{\includegraphics[height=0.90\columnwidth, angle=-90]{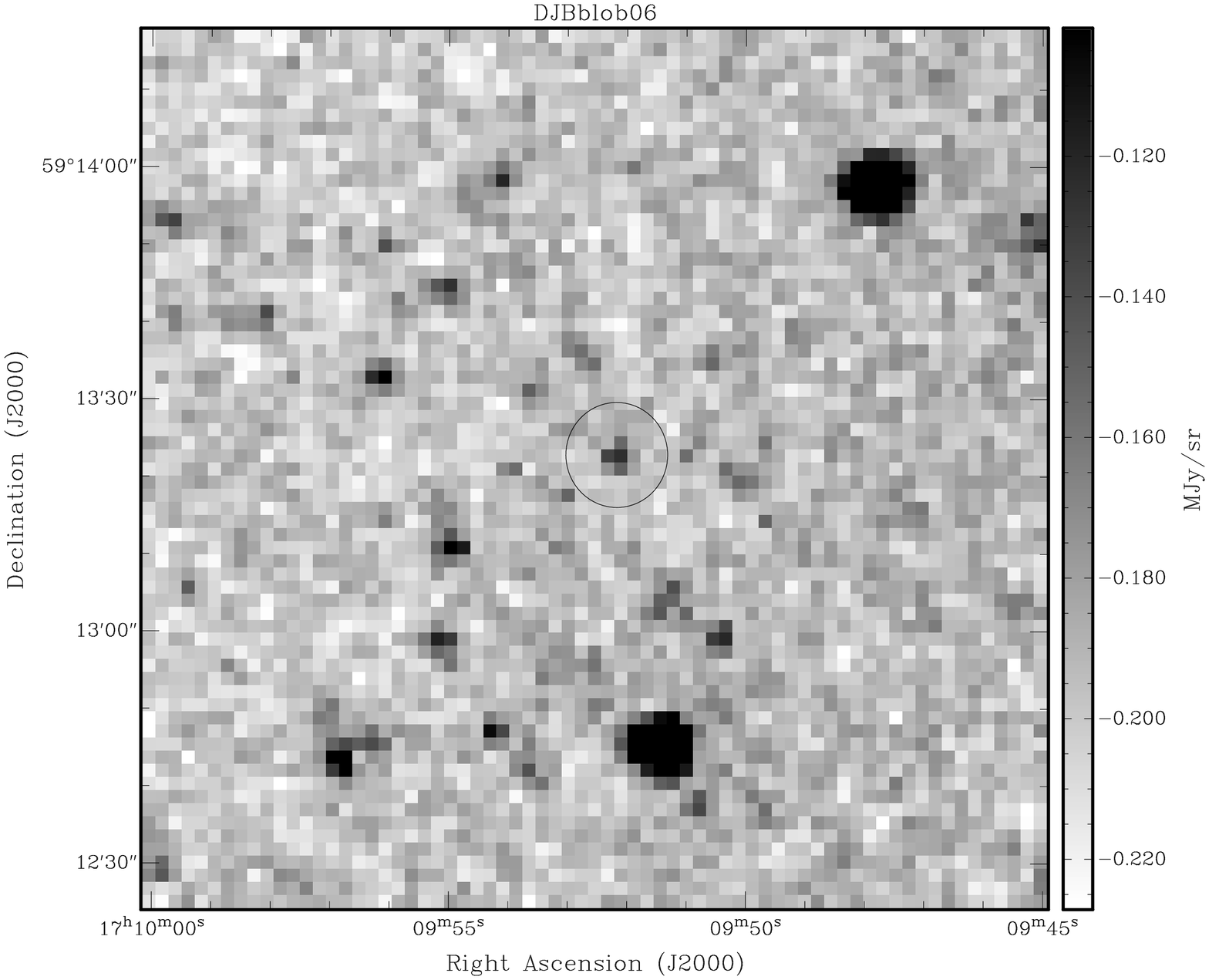}}
\caption[IRAC observations of the LAB counterpart galaxy.]{IRAC
  observations of the LAB counterpart galaxy. The galaxy is centred in
  each frame and identified by the 12\asec\ diameter black circle. The
  counterpart galaxy is clearly detected in each band; for fluxes and
  errors, see table \ref{photometry}.}
\label{IRACch1234}
\end{figure*}

These observations were carried out under program number 40957; the
IRAC data in channels 1 \& 3 were observed on Aug 8$^{\rm th}$, 2007,
with channels 2 \& 4 observed on August 12$^{\rm th}$, and the MIPS
24$\mu$m data observed on August 22$^{\rm nd}$. The raw data were
reduced, and photometric and astrometric solutions were applied
automatically using the Spitzer Science Center pipeline version
S16.1.0. The resulting frames are calibrated to units of surface
brightness (MJy sr$^{-1}$). The reduced IRAC images, centred on the
LAB counterpart galaxy (circled in black) can be seen in figure
\ref{IRACch1234}.

\begin{figure}
\centering
\includegraphics[height=0.90\columnwidth, angle=-90]{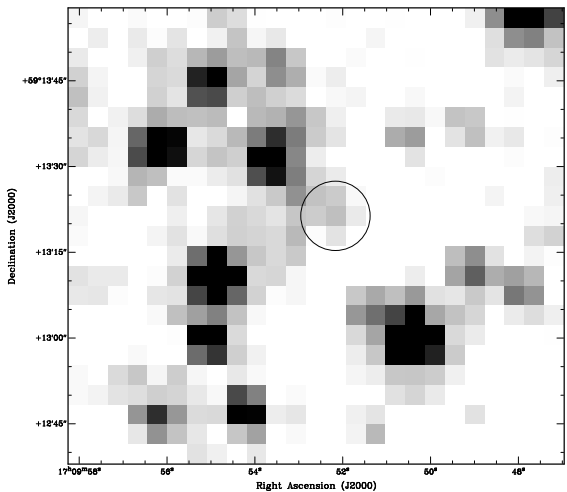}
\caption[MIPS channel 1 observations of the LAB counterpart
  galaxy.]{MIPS channel 1 (24$\mu$m) observations centred on the LAB
  counterpart galaxy (circled). The galaxy is only marginally detected
  in this band (i.e. at $< 3\sigma$); see section
  \ref{sec:Results}. The circle is 12\asec\ in diameter.}
\label{MIPSch1}
\end{figure}

The MIPS pipeline-reduced frames were reduced in seven separate parent
images due to the target jitter pattern employed. They were aligned
according to their astrometric solutions using the {\sc Iraf} task
{\tt wregister}, before being stacked using {\tt imcombine} (also in
{\sc Iraf}). The stacked MIPS data centred on the position of the LAB
(again circled in black), are displayed in figure \ref{MIPSch1}.

To extract meaningful fluxes from the IRAC/MIPS data, a conversion
factor must be applied to convert from MJy sr$^{-1}$ to something more
useful, usually $\mu$Jy pixel$^{-1}$. The conversion factor must take
into account both the difference between MJy and $\mu$Jy, and the
difference between a steradian and the area of each pixel (see
\url{http://ssc.spitzer.caltech.edu/fls/} for further details). Since
Spitzer IRAC imaging data are calibrated to known standard source
observations measured in 10\asec apertures (for MIPS this is 35\asec),
some correction to the photometry is necessary when measuring fluxes
though apertures of different sizes due to the varying amounts of the
point spread function falling within the extent of the aperture. For a
4\asec\ radius aperture in IRAC this factor has values of 1.074,
1.075, 1.080 \& 1.150 for channels 1 to 4 (3.6 - 8.0$\mu$m), whereas
for MIPS channel 1 (24$\mu$m band), the factor is $\sim$2.4, due to
the much larger PSF in MIPS data.

The MIPs $24\mu$m data have a 5$\sigma$ detection limit varying from
$\sim 27.4\mu$Jy at the centre of the field of view to $\sim 58\mu$Jy
towards the edges. We have the sensitivity to detect SFRs as low as
200 \Msolar\ yr$^{-1}$ at 5$\sigma$ in our 24$\mu$m data (although
this does depend on the exact SED assumed, due to the relationship
between 24$\mu$m flux and far infra-red luminosity), and the
accompanying IRAC sensitivity to distinguish between AGN and
starbursting components, were they detected in the 24$\mu$m band.

\subsection{MAMBO-2 Observations}

The Lyman $\alpha$ halo was observed between 19$^{\rm th}$ January and
3$^{\rm rd}$ March 2008 using the Max-Planck Millimetre Bolometer
Array (MAMBO-2, Kreysa et al., 1998) instrument at the IRAM 30m
telescope at Pico de Veleta, Spain. The effective frequency of MAMBO-2
is 250GHz, which corresponds to 1.2mm, with a beam FWHM of
10.7\asec. The observations were carried out in mostly clear
conditions with atmospheric opacity $\tau_{1.2~\rm mm} < 0.3$ and low
sky noise ($<$70~mJy~beam$^{-1}$, with most of the scans having
$\leq$40~mJy~beam$^{-1}$). The targets were centred on the most sensitive
pixel (number 20) and standard on-off observations were carried out
with a wobbler throw in azimuth of 35-45\asec, at a rate of 2~Hz to
enable effective sky subtraction. Individual scans of typically 20
minutes were carried out, and the total integration time was 212
minutes.

To enable effective sky-subtraction, reduce systematic errors and
avoid false positive detections, we varied the wobbler throw between
35, 40 and 45\asec, and scans were conducted at different times on
different days. As a result it is very unlikely for the sky
observations to repeatedly fall on bright nearby sources. Prior to
each scan, pointing corrections were made by observing a nearby bright
source, J1638$+$573. The focus and opacity were checked regularly
(typically every 2~hours) and the focus was generally found to be
stable.  Gain calibration was performed using Neptune, Uranus or Mars,
and monitored regularly using very bright millimetre sources
(typically $\geq$5 Jy); the resulting absolute flux scale has an
uncertainty of $\pm$20\%. The data were reduced by using the MOPSIC
software (Zylka 1998); with these new data we have the sensitivity to
effectively distinguish between the three most plausible sources of
ionization.

\section{Results}
\label{sec:Results}

Due to the crowded nature of the field around the LAB caused by the
depth of the data, 4\asec apertures were preferred over larger options
to prevent the inclusion of flux from neighbouring sources (although
this was not possible for the MAMBO-2 observations, due to the
10.7\asec\ FWHM of the beam). The fluxes were calculated using the
function {\tt APER} from the IDL Astronomy Users'
Library\footnote{\url{http://idlastro.gsfc.nasa.gov/}}, which computes
photometry in specified apertures and background annuli. {\tt APER}
accounts precisely for the different contributions of different square
pixels overlapping on circular apertures, of greater importance here
due to the comparatively large pixel size (1.2\asec). Photometry of
the LAB counterpart is presented in table \ref{photometry}.

\begin{table}
\centering
\caption[Photometric data for the newly-discovered LAB]{Photometric
  data for the LAB. Magnitudes are measured in apertures with 4$\asec$
  radii (with the exception of the 1.2mm data, for which the beam FWHM
  is 10.7\asec), and where they are limits, they are to 1$\sigma$. The
  1.2 mm flux is calculated as detailed in the text, while the radio
  limits are calculated according to the local rms noise in the
  images. The large error in the HeII narrow-band filter has several
  causes; the faintness of the LAB, the difficulty in fitting the sky
  background due to the very extended nature of the LAB, and the
  presence of the nearby low-$z$ interloper to the North and East of
  the LAB counterpart galaxy all contribute. Some values for the
  longer wavelength data are given in units of Jansky (1 Jy =
  10$^{-26}$W~Hz$^{-1}$), which is deemed more appropriate.}
\vspace{0.15cm}
\begin{tabular}{|l|l|}
\hline
Photometric Band &  AB Magnitude\slash Flux  \\ 

\hline
u$^\star$        &  $>$ 26.44                \\
RGO U            &  $>$ 27.85                \\
Harris B         &  23.57$^{\mbox{ \tiny +0.23}} _{\mbox{ \tiny -0.19}}$   \\
Sloan-g'         &  23.97$^{\mbox{ \tiny +0.70}} _{\mbox{ \tiny -0.43}}$   \\
HeII(468.6)      &  22.16$^{\mbox{ \tiny +1.46}} _{\mbox{ \tiny -0.60}}$   \\
Harris V         &  23.97$^{\mbox{ \tiny +0.32}} _{\mbox{ \tiny -0.25}}$   \\
R                &  23.81$^{\mbox{ \tiny +0.20}} _{\mbox{ \tiny -0.17}}$   \\
Sloan-i'         &  24.07$^{\mbox{ \tiny +1.19}} _{\mbox{ \tiny -0.55}}$   \\
J                &  $>$ 19.74                \\
K                &  21.05$^{\mbox{ \tiny +0.20}} _{\mbox{ \tiny -0.18}}$   \\
3.6$\mu$m       & 20.71$^{\mbox{ \tiny +0.04}} _{\mbox{ \tiny -0.04}}$    \\
4.5$\mu$m       & 20.63$^{\mbox{ \tiny +0.05}} _{\mbox{ \tiny -0.04}}$    \\
5.8$\mu$m       & 20.85$^{\mbox{ \tiny +0.36}} _{\mbox{ \tiny -0.27}}$    \\
8.0$\mu$m       & 20.62$^{\mbox{ \tiny +0.24}} _{\mbox{ \tiny -0.19}}$    \\
24$\mu$m        &  18.69 $\pm 7.37 \mu$Jy \\
1.2 mm           &  $< 290 \mu$Jy \\
1.4 GHz          &  $< 20 \mu$Jy \\ 
610 MHz          &  $< 80 \mu$Jy \\ 
\hline
\end{tabular}
\label{photometry}
\end{table}

\subsection{Searching for an obscured AGN within the LAB}

To consider the possibility of an AGN residing within the halo, we
examine the results of our 24$\mu$m channel MIPS observations. Table
\ref{photometry} and figure \ref{MIPSch1} demonstrate the marginal
detection at the position of the LAB in the MIPS 24$\mu$m band of
18.69 $\pm$ 7.37 $\mu$Jy (or in AB magnitudes, m$_{24{\mu m}}$ =
20.72$^{\mbox{ \tiny +0.54}} _{\mbox{ \tiny -0.36}}$). The object is
barely detected, and confusion noise is the dominant source of noise
in the flux estimate. The error in this band was estimated using blank
field aperture measurements of nearby regions of sky with a similar
background of confusing sources. This technique was used to derive the
24$\mu$m photometry in preference to point spread function fitting
since the errors are more easily estimated (in particular, estimating
the true background is almost impossible in these data due to the
effects of confusion). If this marginal detection is real, it may be
indicative of an obscured AGN component or an ongoing starburst
residing within the halo.

To quantify the likelihood that this detection in the 24$\mu$m band
observations is real, we turn to Gaussian statistics. This value can
be obtained by chance approximately 1.1\%\ of the time given the error
associated with the measurement, however this value must be treated as
a lower limit due to the presence of nearby sources whose point spread
functions overlap with the 4\asec\ flux extraction aperture used for
this measurement.

\begin{figure*}
\centering \includegraphics[width=2.00\columnwidth]{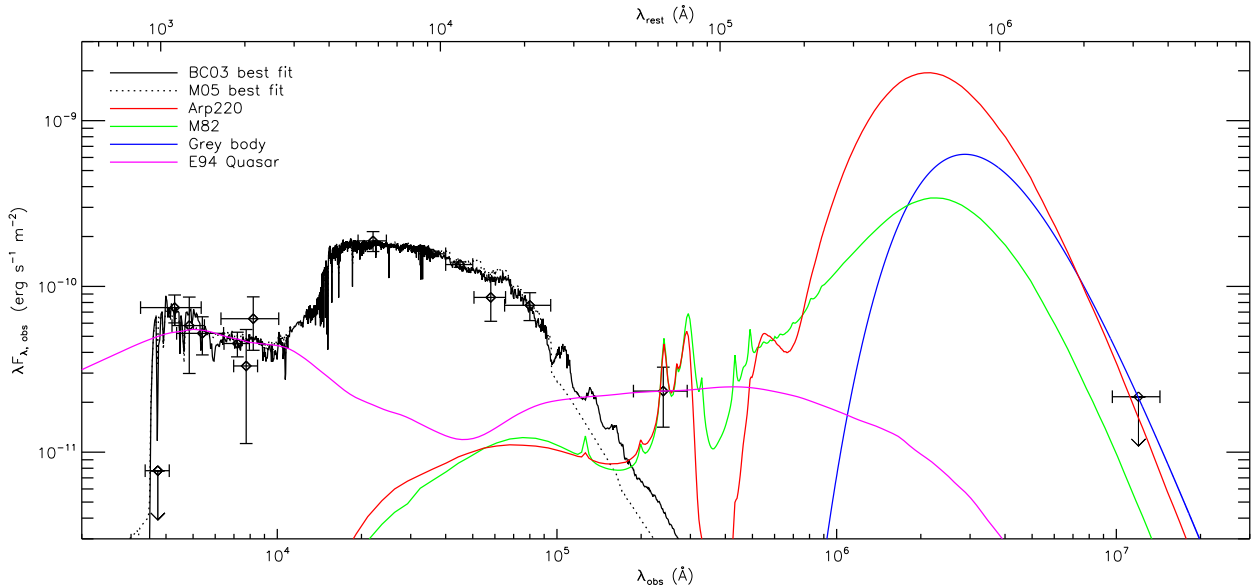}
\caption{Multi-wavelength spectral energy distribution of the LAB's
  counterpart galaxy. The best fit models to our photometry from the
  Bruzual-Charlot \& Maraston simple stellar population models (solid
  and dotted black lines, respectively). Also included are Arp220, M82
  and template QSO (from Elvis et al. 1994) SEDs shown in red, green,
  \& purple respectively, and the best-fit grey-body emission profile
  (blue). The Bruzual \& Charlot (2003) and Maraston (2005) SEDs are
  normalised to the 3.6$\mu$m band, the Arp220, M82 and QSO templates
  are normalised to the 24$\mu$m band, and the greybody is normalised
  to the 1.2mm MAMBO-2 observations. We find that our optical \& IRAC
  data are very well described by a composite simple stellar
  population consisting of a young and an old component aged 10 Myr
  and 0.8Gyr for the Bruzual \& Charlot (2003) models, or 10 Myr and
  2.0Gyr for the Maraston (2005) models. The weak detection at
  24$\mu$m, which is questionable due to its low signal-to-noise as
  well as the approaching confusion limit, indicates either a weak
  obscured AGN component or a starburst. Our other data strongly
  constrain these possibilities -- see text. While the unobscured
  template quasar SED is consistent with our optical photometry, it is
  inconsistent with the results of our spectroscopy, which do not
  detect the presence of the high ionisation emission lines that the
  presence of such a naked AGN would suggest. }
\label{SED_IRACnorm}
\end{figure*}

We adopt the conservative assumption that the 24$\mu$m detection is
real in order to estimate the possibility of an AGN or a starburst
being the power source of the extended Lyman-$\alpha$ emission. In
this way, we can estimate a limiting luminosity for any AGN residing
in the LAB from the 24$\mu$m flux density, which is not thought to be
affected by dust obscuration. If this marginal detection were entirely
due to an AGN, the monochromatic luminosity would correspond to a
rest-frame luminosity of $L_{\nu_{\rm RF}}= 3\times
10^{23}$~W~Hz$^{-1}$ at $\lambda_{\rm RF}$~=~6.3~$\mu$m. Assuming the
Elvis et al. (1994) quasar SED we estimate that this corresponds to
$L_{\rm bol}=3.4\times10^{11}$~L$_{\odot}$ ($1.3\times10^{45}$ erg
s$^{-1}$), of which $\sim 8.6 \times 10^{43}$ erg s$^{-1}$ is emitted
at wavelengths between 200-912\AA\ in the rest frame, and thus capable
of powering Lyman-$\alpha$ emission (following Dey et al., 2005, and
assuming case B recombination).

At first glance, it seems plausible that an unobscured AGN is
energetically capable of powering a reasonable proportion ($\sim30$\%)
of the 2.83 $\pm\ 0.5 \times 10^{44}$ erg s$^{-1}$ Lyman-$\alpha$
emission that we observe. However, we can also use a photon number
counts argument similar to that in Neugebauer et al. (1980) to test
the possibility of the halo being ionised by an AGN. The
Lyman-$\alpha$ luminosity of the halo requires a total of $1.7 \times
10^{55}$ ionising photons per second. Integrating under the Elvis et
al. (1994) quasar SED between 200 and 912\AA\ (it is only these
photons which contribute to the ionisation of neutral Hydrogen, and
thus plausibly power the Lyman-$\alpha$ emission) suggests that any
putative quasar component may emit only 2.3$\times 10^{54}$ photons
each second. This figure is almost an order of magnitude fewer than
that required to result in the Lyman-$\alpha$ luminosity that we
observe. It seems improbable that an AGN is powering the LAB; the
quasar SED normalised to our 24$\mu$m detection is displayed in figure
\ref{SED_IRACnorm}.

Furthermore, it should be noted that LABs containing AGN (e.g. that
which was presented in Dey \etal\ 2005) and the similarly luminous
Lyman-$\alpha$ haloes belonging to some high redshift radio galaxies
at $z > 2$ (e.g. Villar-Martin 2007a, compared with 24$\mu$m data from
Seymour et al. 2007) are detected at mid-infrared wavelengths at
fluxes typically at least an order of magnitude more luminous than
those probed by our 24$\mu$m observations.  Our view that an AGN is
not responsible for the Lyman-$\alpha$ emission observed here is
further supported by the lack of other evidence for an AGN at radio
wavelengths in the GMRT and VLA data (at 610 MHz and 1.4GHz
respectively), or in our optical spectroscopy, where no other emission
lines were detected (see Smith \& Jarvis, 2007).

\subsection{Constraining the star formation rate with MAMBO-2 and MIPS}

To probe for evidence of the obscured star formation rates implied by
superwind models of LABs of order $\sim$1000\Msolar~yr$^{-1}$
(e.g. Taniguchi \& Shioya, 2000), we turn first to the results of our
MAMBO-2 observations. The host galaxy is not detected at 1.2~mm down
to a 3$\sigma$ flux limit of 0.86~mJy. To constrain the star-formation
rate probed due to cold dust masses within this halo, we consider the
1.2mm (observed frame) flux to be due to a {\it grey body}, with
emissivity of the form given in equation \ref{greybody} which is also
displayed in figure \ref{SED_IRACnorm};

\begin{equation}
F_\lambda \propto \frac{1}{\lambda} ~ \frac{ \frac{1}{\lambda}^{(4+\beta)} }{ e^{(\frac{hc}{\lambda kT})} - 1 },
\label{greybody}
\end{equation}

\noindent where $\beta$\ corresponds to the emissivity index of the
grey body spectrum and T to the temperature of the dust. We assume
typical values for these two parameters from Omont \etal\ (2003) of
$\beta$ = 1.5 and T=45K, and adopt a conversion from far-infrared
luminosity between 8 \& 1000$\mu$m and star formation rate (SFR)
following Kennicutt \etal\ (1998). With these assumptions we find that
our 3$\sigma$ flux limit at 1.2mm suggests that the star formation
rate within the LAB is $<$550\Msolar ~yr$^{-1}$. This puts our three
sigma upper limit for the star formation rate within this LAB slightly
below the SFRs required by superwind models of LAB ionization. If we
adopt a slightly lower temperature for the dust component, T=35K, our
estimate for the SFR within the halo becomes $<$220\Msolar yr$^{-1}$
to 3$\sigma$.

We can also use our MIPS observations to constrain the SFR. To obtain
SFR estimates from 24$\mu$m fluxes, one must assume a particular SED
in order to enable an estimate of the far infra-red luminosity, and
then proceed as for the {\it grey-body} models. Instead of assuming a
{\it grey-body} SED, here we use templates for M82 and Arp220 from
Siebenmorgen \& Kr\"ugel (2007) to derive SFR estimates. If our
detection at 24$\mu$m is real, then it corresponds to SFRs of $\sim$
140 (M82 template) or 620 (Arp220 template) \Msolar yr$^{-1}$. Given
the uncertainties in these estimates, both values are consistent with
the limit derived from our MAMBO-2 observations.

\subsection{The Stellar Population}

Having essentially ruled out the presence of a powerful AGN, we turn
our attention to the stellar population of the LAB. The new IRAC
photometry of the LAB counterpart galaxy provides much better
constraints for its stellar SED, enabling us to calculate a much more
effectively constrained best-fit SED, which is displayed in figure
\ref{SED_IRACnorm}. The better constraints on the SED morphology then
enable more accurate estimates of the stellar mass of the counterpart
galaxy.

In Smith \& Jarvis (2007), uncertainties in the data meant that the
choice of template SEDs used for the SED fitting was not crucial;
using either the Maraston (2005) or Bruzual \& Charlot (2003) template
SEDs made little difference to the outcome. This is no longer the case
due to our new infrared and millimetre-wavelength data, combined with
deeper optical data obtained from the William Herschel Telescope's
prime focus imager (PFIP) to investigate the environment of the LAB
(Smith \& Jarvis {\it in prep}). The difference between the two models
(caused by different treatments of thermally pulsing asymptotic giant
branch -- TP-AGB -- stars) at rest-frame near-infrared wavelengths
becomes important due to the quality of our IRAC observations. The
Maraston models are generally redder, and the effects of the TP-AGP
manifest themselves as increased flux at near infrared wavelengths
(e.g. Maraston, 2005). Since the models vary, we used both sets to
approximate our SED. In order to enable reasonable comparisons between
the best-fit models of each type, we required that both fits made use
of the Salpeter (1955) initial mass functions.

The best fit SED was calculated for both species of SSP models with
ages less than the 3 Gyr (the age of the Universe at $z = 2.83$ in our
adopted cosmology is 2.344 Gyr -- the reason for this range of SSP
ages will become clear), using a simple $\chi^2$ fitting algorithm. We
normalised the composite SED to the IRAC 3.6$\mu$m data point (channel
1), since it has the smallest photometric errors of all bands, and is
less susceptible to the effects of dust than observations in any of
the shorter-wavelength bands. A variety of composite stellar SEDs were
fit to our LAB counterpart photometry (table \ref{photometry}),
consisting of linear combinations of young ($< 600$Myr) and old ($>
600$Myr) simple stellar populations. The best fit models based on the
Bruzual-Charlot and Maraston SSPs are displayed in figure
\ref{SED_IRACnorm}.

Whilst the properties of the two fits appear similar, the underlying
properties are significantly different. The best-fit Bruzual \&
Charlot (2003) composite fit consists of a linear combination of
components aged 10 Myr and 0.8 Gyr, with Salpeter IMFs (Salpeter,
1955) and solar metallicity. The best fit Maraston (2005) model
comprises components aged 10 Myr \& 2.0Gyr with Salpeter IMFs, and low
metallicity (Z = 0.0001 = $\nicefrac{1}{50}~Z_{\odot}$). The two
best-fit models differ widely in their properties in all but the
broadest sense (that the best fit consists of a composite simple
stellar population with both young and old components).

Whilst the metallicities appear widely different, the discrepancy in
values of $\chi^2$ parameter suggests that our data are not capable of
effectively distinguishing between different values of $Z$. Whilst the
best-fit Maraston (2005) model has $Z = 0.0001$ and $\chi^2 = 10.16$,
if we set $Z = 0.02 = Z_\odot$ then the best-fit model has $\chi^2 =
13.19$ (albeit with a different combination of SSPs -- 10 Myr \& 3.0
Gyr, older than the age of the Universe at $z = 2.83$). With the data
available to us, attempts to effectively constrain values of the host
galaxy's metallicity are uncertain.

The masses of each component were derived from the normalisations of
the models used in the $\chi^2$ fitting process. The mass of the
Bruzual \& Charlot (2003) best-fit model is 3.42 $\pm~0.13 \times
10^{11}$\Msolar, while for the Maraston (2005) models the total mass
is 4.36 $\pm~0.16 \times 10^{11}$\Msolar. Here, the errors are derived
from the normalisation based upon the IRAC 3.6$\mu$m flux, and are
probably underestimates as a result. The details of each best-fit
population, including masses, ages and errors are shown in table
\ref{bestfit}. 

\begin{table}
\centering
\caption{Properties of the best fit composite simple stellar
  populations for both the Bruzual \& Charlot (2003) and Maraston
  (2005) models. The age, masses and errors are presented for both of
  the simple stellar populations that best fit our extensive
  photometry.}
\vspace{0.15cm}
\begin{tabular}{l|l|l}
\hline
Parameter & Bruzual \& Charlot (2003) & Maraston (2005) \\
\hline
Age$_1$(Myr) & 10 & 10 \\
Mass$_1$(\Msolar) & $9.06 \pm 0.34 \times 10^8$ & $5.88 \pm 0.22 \times 10^8$ \\
\hline
Age$_2$(Gyr) & 0.8  & 2.0  \\
Mass$_2$(\Msolar) & $3.41 \pm 0.13 \times 10^{11}$ & $4.35 \pm 0.16\times 10^{11}$ \\
\hline
Mass$_{\rm total}$(\Msolar) & $3.42 \pm 0.13 \times 10^{11}$ & $4.36 \pm 0.16 \times 10^{11}$ \\
Best $\chi^2$ & 7.24 & 10.16 \\
\hline
\end{tabular}
\label{bestfit}
\end{table}

Since the old SSP in the Maraston model fitting scheme is older (2
Gyr) than the old SSP used in the best fit in the Bruzual-Charlot
scheme, one would expect it to be more quiescent. That being the case,
a larger stellar mass would be required to produce the same luminosity
due to the death of the more luminous (and {\it shorter-lived})
stars. The different mass estimates from the Bruzual \& Charlot (2003)
and Maraston (2005) schemes are not unexpected, since one of the main
disagreements between these two models of SEDs relates to the stellar
masses implied (e.g. Maraston, 2005).

\begin{figure}
\centering
\includegraphics[width=0.90\columnwidth]{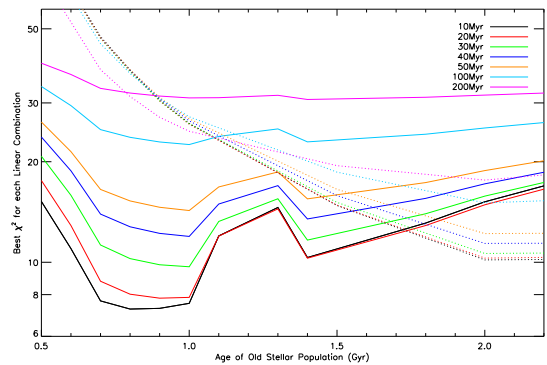}
\caption{The distribution of best-fit values of $\chi^2$ for composite
  Bruzual \& Charlot (2003) and Maraston (2005) simple stellar
  populations (solid and dotted lines, respectively). The age of the
  best-fit old SSP component is shown on the abscissa, with the age of
  the best-fit young SSP component represented by the colours (see
  legend). }
\label{chi_dist_IRACnorm}
\end{figure} 

The variation of best-fit $\chi^2$ as a function of input models is
displayed in figure \ref{chi_dist_IRACnorm} for the two fitting
regimes. A clear minimum occurs for the combination of 10 Myr \& 0.8
Gyr Bruzual \& Charlot templates. The preference for this model over
other similar combinations is small, but this combination is the best
fit, in agreement with the results of Smith \& Jarvis (2007), which
suggested that the counterpart galaxy was best described by a
combination of a young and an old simple stellar population. The
Bruzual \& Charlot models (2003 - solid lines) better approximate our
data than those of Maraston (2005 - dotted lines), reflected by the
lower values of $\chi^2$. Maraston models older than the age of the
Universe are included in this analysis due to the coarse sampling of
population ages above 1.5 Gyr, to enable the models to demonstrate a
minimum $\chi^2$ value. Thus we conclude that the host galaxy is
well-described by a composite simple stellar population, with little
or no evidence for additional (e.g. AGN) components. We also note that
the composite stellar population is considerably (i.e. orders of
magnitude) less luminous between 200 \& 912\AA\ than the Elvis (1994)
AGN SED, and therefore also incapable of powering the Lyman-$\alpha$
halo.

\section{Conclusions}
\label{conclusions}

With our new IRAC, MIPs and MAMBO-2 data, as well as our new deep U, B
\& V band observations, we have demonstrated that the counterpart
galaxy associated with the newly-discoverd LAB announced in Smith \&
Jarvis (2007) is well-described by a composite stellar
population. Using this new multi-wavelength data-set, we find no
plausible evidence for the presence either of an obscured active
galactic nucleus or of an energetic starburst that would be required
to ionize the neutral Hydrogen gas in galaxy-wide super-wind
schemes. The most likely source of ionization for this particular
highly luminous Lyman-$\alpha$ halo is a ``cold accretion'' scenario,
which may be able to supply sufficient energy for this profuse
emission without invoking either of the other power sources suggested
in the literature (see Fardal \etal\ 2001, Dijkstra, \etal, 2006a,b,
2007). We also place further constraints on the host galaxy, and find
it to be well described by composite simple stellar populations with
total masses of 3.42 $\pm~0.13 \times 10^{10}$ or 4.36 $\pm~0.16 \times
10^{11}$\Msolar, depending on the models used.

\section*{Acknowledgments}
DJBS wishes to thank Chris Simpson for valuable conversations, and the
UK STFC for a PDRA. MJJ acknowledges the support of a Research Council
UK fellowship. This work is based in part on observations made with
the {\it Spitzer Space Telescope}, which is operated by the Jet
Propulsion Laboratory, California Institute of Technology under a
contract with NASA. Support for this work was provided by NASA. Many
thanks to the IRAM staff for their support, particularly St\'ephane
Leon for running the MAMBO pool, and to all guest observers during the
pool observing sessions at the 30m. IRAM is supported by INSU/CNRS
(France), MPG (Germany), and IGN (Spain). The Isaac Newton and William
Herschel Telescopes are operated on the island of La Palma by the
Isaac Newton Group in the Spanish Observatorio del Roque de los
Muchachos of the Instituto de Astrofisica de Canarias.

\bsp

\label{lastpage}

\end{document}